\documentclass[vecphys]{svmult}

\usepackage{makeidx}         
\usepackage{graphicx}        
\usepackage{multicol}        
\usepackage[bottom]{footmisc}

\newcommand{\arcmin}{\mbox{$^\prime$}}%
\newcommand{\arcsec}{\mbox{$^{\prime\prime}$}}%
\newcommand\micron{\mbox{$\mu$m}}%

\makeindex             

\begin{document}

\title*{TMT Science and Instruments}
\author{David Crampton, Luc Simard \and David Silva}
\institute{TMT Project Office, 2636 East Washington Blvd, Pasadena, CA 91107, USA
\texttt{crampton@tmt.org}}
%
%
\maketitle

\paragraph{Abstract}

To meet the scientific goals of the Thirty Meter Telescope Project, full diffraction-limited performance is required from the outset and hence the entire observatory is being designed, as a system, to achieve this. The preliminary design phases of the telescope and the first light adaptive optic facility are now approaching completion so that much better predictions of the system performance are possible. The telescope design and instrumentation are summarized in this presentation, with a brief description of some of the scientific programs that are foreseen.




\section{Introduction}
\label{sec:1}
The TMT Project is rapidly moving towards construction of a 30m telescope that is being designed from the outset as a system that will deliver diffraction limited images at wavelengths longer than 1 micron. Many science programs will thus realize the D$^4$ advantage in point source sensitivity inherent in such a telescope. Although a primary diameter, D, greater than 30m would offer an even larger gain, our analyses indicate that 30m is the optimum balance between science benefit, cost, technological readiness and schedule at the present time. Using components that are currently, or soon-to-be available, it is possible to achieve images with a high Strehl ratio at wavelengths greater than $1\micron$ with a 30m telescope and, although the instruments are challenging, they are feasible. The telescope will be a Ritchey-Chretien design with a f/1 primary mirror. The latter will be composed of 492 1.4m segments. Instruments will be located on two large Nasmyth platforms, addressed by an articulated tertiary mirror. This will enable rapid (less than 10 minutes) switching between on-sky observations of targets with different instruments (less than 5 minutes with the same instrument). Adaptive optic (AO) systems, including a laser guide star facility, are being integrated into the observatory system, with plans to employ multi-conjugate AO (MCAO), multi-object AO (MOAO), ground layer AO (GLAO), mid-IR AO (MIRAO) and ``extreme'' AO (ExAO). The telescope will produce a $20 \arcmin$ diameter field for seeing limited observations from the ultraviolet ($0.3\micron$) to the MIR ($28\micron$). Aerodynamic studies demonstrate excellent performance of the Calotte style enclosure, providing venting to minimize image degradation within the enclosure while not inducing unacceptable wind buffeting of the telescope itself.

The design and development phase of the TMT project will reach completion in early 2009. Science operations are expected to begin in the last half of the next decade.  

\section{Instrument Suite}
\label{sec:2}

The TMT Scientific Advisory Committee (SAC) identified a comprehensive suite of eight instruments required to tackle the science that they envisage for the first decade of operation.  The proposed instruments span the discovery space in wavelength, spatial resolution, spectral resolution (R) and field-of-view/slit length. They also define a number of important TMT subsystem parameters such as the  physical sizes and weights of instruments that the observatory should be designed to accommodate. Six of the instruments use built-in AO systems or use NFIRAOS, the facility MCAO system, to exploit the diffraction-limited capability of TMT. The other two are seeing-limited but could utilize AO  to improve their observing efficiency. The instruments also exploit the entire wavelength range of TMT, from 0.31 to $28 \micron$; they include a high contrast instrument; and instruments with a wide variety of field sizes, up to $20 \arcmin$ in diameter. Thus the instrument suite is representative and suitable for defining general instrument requirements that should provide enough flexibility to accommodate future instruments.
Feasibility studies of the instruments were carried out in 2005-2006. Nearly two hundred scientists and engineers at forty-six US, Canadian and French institutions were involved in these studies, which were reviewed by panels of international experts. Most importantly, perhaps, at this stage, these studies demonstrate that the instruments are feasible, albeit challenging. The science cases and operational concept documents of these studies highlight and document the tremendous scientific potential of TMT.

A diagram demonstrating how the instruments could be arranged on the Nasmyth platforms is shown in Figure 1.

\begin{figure}[h] 
   \centering
   \includegraphics[width=4in]{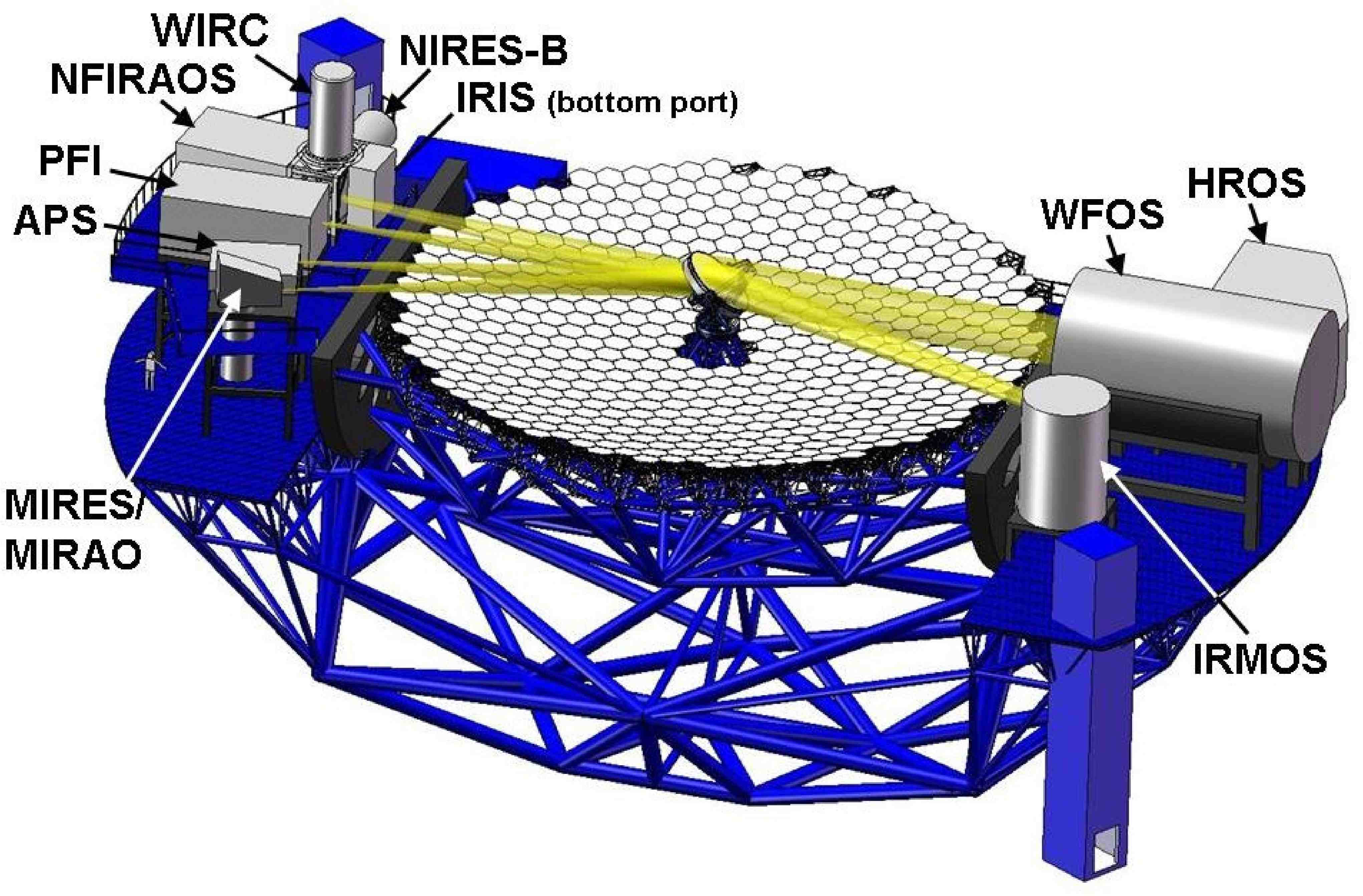} 
   \caption{Diagram showing how the instruments, located on two large Nasmyth platforms, will be addressed by the articulated tertiary mirror. See text for meaning of the acronyms. IRMS will be located on the side port of NFIRAOS (where NIRES-B is shown) during the first years of operation. The ``instrument"  labeled APS is the system used to align and phase the 492 segments of the primary mirror and the optical system as a whole.}
   \label{fig:FD}
\end{figure}

\subsection{Early Light Instruments}
\label{sec:3}
The instrument suite has been divided by the SAC into ``early light'' and ``first decade'' instruments for a variety of pragmatic reasons, mostly to do with funding constraints, commissioning practicalities, and technological readiness. The early light suite consists of IRIS, a NIR integral field spectrograph and imager working at the diffraction limit, WFOS, a widefield multiobject spectrograph, and IRMS, a multislit NIR spectrograph and imager fed by the facility MCAO system. While bringing the early light instruments on-line is clearly the top priority of the TMT instrumentation program,  the ultimate goal is to bring the full suite into operation over the first decade of  operations. Many global observatory design decisions and choices were therefore made with this in mind. The very ambitious WFOS originally requested by SAC has been scaled back to reduce cost, risk and commissioning complexity and make it suitable for early light. Likewise, the early light IRIS configuration has also been kept as simple as possible yet still retaining the ability to meet its key science objectives.

\paragraph{InfraRed Imaging Spectrograph (IRIS)}
 
 IRIS, located on the bottom port of NFIRAOS,  will be able to conduct diffraction-limited imaging and integral field spectroscopic observations in the 0.8-2.5$ \micron$ wavelength region. The SAC has consistently ranked IRIS as the top instrument priority for TMT, partly because of its ability to utilize the exquisite images delivered by NFIRAOS (which will be only 7 milliarcsec in the J band).  The imager will have a field of view of at least $15\arcsec$ and IRIS is expected to incorporate several different plate scales for integral field spectroscopy yielding fields of view up to $2\arcsec$ . Astrometric measurements with precisions of order 100 microarcsec are foreseen. 
 
 The versatility of IRIS allows it to address a very broad range of science problems. Some of the prime science drivers include observations of  the first luminous objects in the Universe, supermassive black holes in the cores of distant galaxies, relativistic effects at the Galactic Center, resolved stellar populations in the crowded fields of galaxies out to the Virgo galaxy cluster, and high-resolution imaging and spectroscopy of planets and satellites in the Solar System.

\paragraph{InfraRed Multi-Slit Spectrograph (IRMS)}
Some form of multi-object NIR spectroscopy is another essential capability for early light. Understanding the so-called ÒFirst LightÓ objects in the Universe, the origin and evolution of galaxies and other objects detected by JWST and ALMA will require spectra of many extremely faint objects in the NIR, and multiplexing will thus be essential. Although a fully multiplexed deployable IFU system using MOAO was judged to be too risky and expensive for an early light instrument, fortuitously a clone of the MOSFIRE multislit instrument, currently being built for Keck, provides a very exciting interim capability. Although MOSFIRE will be a seeing-limited instrument for Keck, it can be easily adapted for use in an AO mode with NFIRAOS, providing an exceedingly powerful capability for TMT at low risk and modest cost. When optimized for widefield mode, NFIRAOS will deliver images to IRMS that will produce almost an order of magnitude gain in encircled energy within narrow (160mas) slits over the entire of $2\arcmin $ diameter field.

\paragraph{Wide-Field Optical Spectrograph (WFOS)}

A number of key TMT science programs will be best conducted with a powerful survey instrument operating at optical wavelengths. These programs include the determination of the baryonic power spectrum, the tomography of the intergalactic medium, the determination of the dynamical states of stellar populations in nearby galaxies, the dark matter distribution in elliptical galaxies, and the star formation history in local galaxies. A high multiplexing capability (several hundreds of spectra) over a relatively large field-of-view is required to sample large cosmological volumes with sufficient target density to probe the range of desired physical scales. The spectrograph should also provide good image quality and moderate spectral resolutions over the wavelength range $0.31-1.1\micron$.

\subsection{First Decade Instruments}
\label{sec:4}
\paragraph{InfraRed Multiple Object Spectrometer (IRMOS)}
IRMOS, as envisaged by SAC, is simultaneously perhaps the most ambitious and the most exciting of the TMT instruments. Its goal is to deliver 2D integral field spectroscopy of many objects over a 5\arcmin\ field of regard using MOAO to deliver quasi-diffraction-limited spatial resolution in the NIR. Key IRMOS science includes the physical properties of galaxies (internal velocity fields, star formation rate, chemical abundance) at the epoch of peak galaxy assembly ($z \sim 2-3$) and the physical conditions in star-forming regions. IRMOS will allow many outflows around newly-forming stars to be resolved and their interaction with the interstellar medium to be studied. 

\paragraph{Near-Infrared Echelle Spectrometer (NIRES)}
NIRES is a straightforward NIR echelle spectrograph for the $1-2.4\micron$ wavelength range that also  fits behind NFIRAOS. In fact, NIRES could effectively be a clone of existing or planned diffraction-limited NIR spectrographs (e.g. Keck NIRSPEC). The NIRES feasibility study clearly identifies the enormous potential of such an instrument on TMT for programs such as understanding the physics of gamma-ray bursters, probing the intergalactic medium at high redshifts, and delivering precision radial velocities of late type stars for planet searches. 

\paragraph{High Resolution Optical Spectrometer (HROS)}
High-resolution optical spectrographs have occupied center stage in recent years thanks to a range of exciting work: the very productive Doppler searches for exoplanets, the measurements of chemical abundances in absorbing intergalactic matter along the lines-of-sight to distant quasars and the surveys of the outer reaches of the Milky Way in search of the most metal-poor stars. HROS is fundamentally a seeing-limited, high-resolution optical spectrometer, although it is recognized that its performance could be enhanced by the use of LTAO (Laser Tomography AO). HROS will provide a spectral resolution of $R = 50000$ (1 arcsec slit) or $R \ge 90000$ (image slicer) over the wavelength range: $0.31-1.0 \micron$.

\paragraph{Mid-Infrared Echelle Spectrometer (MIRES)}
MIRES brings diffraction-limited, high spatial-resolution imaging and high-resolution spectroscopy in the thermal infrared (5-25\micron) to the TMT instrumentation suite. The MIRES science case features many fascinating objectives: the origin of stellar mass, the exploration of the inner parts of protoplanetary disks, astrochemistry, and the deposition of pre-biotic molecules onto planetary surfaces. The spectral resolution of MIRES will be $5000< R <100000$ with a diffraction-limited slit.

\paragraph{Planet Formation Instrument (PFI)}
The Planet Formation Instrument (PFI) is focused on the direct detection and characterization of extrasolar planets. PFI is unique among the instruments in that it places significant requirements on the telescope optics (primary pupil shape, secondary support structure, segment edge and reflectivity) and vibrational environment, and these requirements have been factored in the telescope design. PFI will build very strongly on the heritage being gained by the ``planet finder'' instruments that are currently being designed for Gemini and the VLT. The contrast requirements are $10^{-8}$ at 50 milliarcsec, goal of  $10^{-9}$ at 100 milliarcsec (parent star magnitude $I < 8$).

\section{TMT Science}
\label{sec:5}

It is obviously extremely difficult to foresee what scientific programs will be carried out by TMT and many of the discoveries made will be unanticipated or even unimagined at present. However, the SAC has provided examples of potential programs to guide the development of the observatory.  Some of these were mentioned above in the descriptions of each instrument but, in practice, several instruments and attributes of the Observatory will contribute to solving most of the major questions. Two examples of this, and how TMT might be used to address them, are summarized here.

\paragraph{Fundamental Physics and Cosmology}

Various probes of Dark Energy will be undertaken by TMT, including precise measurements of the expansion history and power spectrum of the Universe at low and high redshifts using supernovae. The latter will also be used along with Gamma Ray Bursters (GRB) and Super Massive Black Holes to study the physics of extreme objects. Problems such as determining whether there are variations in the fundamental physical constants require high spectral resolution observations with signal-to-noise that only an extremely large telescope can provide. The nature of Dark Matter will be examined using probes such as the determination of 3D orbits of stars near the Galactic Center, measurement of the baryonic power spectrum through tomography of the InterGalactic Medium (IGM),  and the determination of the kinematics of stars in dwarf galaxies using precision spectroscopy with WFOS, HROS, NIRES and IRIS, and precision astrometry with  IRIS and WIRC. For studies of the sources of first light and cosmic reionization, TMT will have strong synergy with JWST and future 21cm surveys. Although it is anticipated that JWST will be able to detect the brightest such sources, TMT should go at least one magnitude fainter and perhaps much more, depending on their size. At slightly lower redshift, IRIS, IRMOS and NIRES will study detailed properties of first galaxies and their influence on the IGM. WFOS and IRMS spectra of large samples of distant galaxies and the intervening IGM, coupled with 3D spectroscopy of samples of galaxies with IRMOS will elucidate how galaxies acquire gas, how star formation proceeds, and the effects of supernovae, SMBH and active star formation on the formation of galaxies.

Some of these projects gain not only through the huge sensitivity gain provided by the diffraction limited images delivered by TMT but also because study of fields in the galactic centre and nearby galaxies are fundamentally limited by confusion (crowding). In addition, all projects will gain from the rapid target acquisition and rapid switching between instruments that is being designed into the system. Observation of GRBs, especially, will benefit from the fact that TMT will rapidly slew and acquire targets, set up active and adaptive optics systems and be ready to begin observation with any instrument in less than 10 minutes.


\paragraph{Formation of Stars and Planets}

HROS and NIRES will be used to detect and measure orbits and masses of exoplanets. PFI will be able to directly image and characterize a complementary sample of planets. MIRES, using the high angular resolution and high sensitivity of TMT,  will be able to study the atmospheres of planets and protoplanetary disks. These studies will help answer questions related to our place in the Universe:  how and when planets are formed and how might life arise on such planets.

\section{Summary}

TMT will provide a major advance in mankind's ability to probe a very broad range of physics of the Universe. Partial construction funding has just (Dec 5, 2007) been announced by the Gordon and Betty Moore Foundation, ensuring that the project will be able to transition smoothly into the construction phase. 

Many more examples and details of programs that will be enabled by TMT are discussed in the {\it{Detailed Science Case}}  which can be found at www.tmt.org (see Foundation Documents) along with several other documents and references that describe the Observatory in much more detail than is possible here.

\paragraph{Acknowledgments}

The TMT Project gratefully acknowledges the support of the TMT partner institutions. They are the Association of Canadian Universities for Research in Astronomy (ACURA), the California Institute of Technology and the University of California. This work was supported as well by the Gordon and Betty Moore Foundation, the Canada Foundation for Innovation, the Ontario Ministry of Research and Innovation, the National Research Council of Canada, the Natural Sciences and Engineering Research Council of Canada, the British Columbia Knowledge Development Fund, the Association of Universities for Research in Astronomy (AURA) and the U.S. National Science Foundation. 
%
%
%
%


\printindex
\end{document}